\documentclass[12pt]{article}
\usepackage{amsmath}
\usepackage{cite}
\begin{document}
\begin{flushright}
KANAZAWA-17-06\\
September, 2017
\end{flushright}
\vspace*{1cm}

\renewcommand\thefootnote{\fnsymbol{footnote}}
\begin{center} 
{\Large\bf Dark matter stability and one-loop neutrino mass 
generation based on Peccei-Quinn symmetry}
\vspace*{1cm}

{\Large Daijiro Suematsu}\footnote[1]{e-mail:
~suematsu@hep.s.kanazawa-u.ac.jp}
\vspace*{0.5cm}\\

{\it Institute for Theoretical Physics, Kanazawa University, 
Kanazawa 920-1192, Japan}
\end{center}
\vspace*{1.5cm} 

\noindent
{\Large\bf Abstract}\\
We propose a model which is a simple extension of the KSVZ invisible axion 
model with an inert doublet scalar.
Peccei-Quinn symmetry forbids tree-level neutrino mass generation and 
its remnant $Z_2$ symmetry guarantees dark matter stability.
The neutrino masses are generated by one-loop effects as a result
of the breaking of Peccei-Quinn symmetry through a nonrenormalizable 
interaction. Although the low energy effective model coincides with 
an original scotogenic model which contains right-handed neutrinos with 
large masses, it is free from the strong $CP$ problem. 
 
\newpage
\setcounter{footnote}{0}
\renewcommand\thefootnote{\alph{footnote}}

\section{Introduction}
The standard model (SM) has been confirmed by the discovery of the 
Higgs scalar \cite{higgs}. However, it is now considered to be extended 
to explain several experimental and observational data such 
as neutrino masses and mixings \cite{nexp,t13}, and dark matter (DM) \cite{dm}. 
Strong $CP$ problem is also one of such problems suggested by an experimental 
bound of the electric dipole moment of a neutron \cite{strongcp}. 
Invisible axion models are known to give a simple and interesting solution 
to it \cite{ksvz,dfsz}. 
The KSVZ model, which is one of such realizations, is an 
extension of the SM by a complex singlet scalar and a pair of colored 
fermions. 
It has a global $U(1)$ symmetry, which is violated only by the
QCD anomaly and plays a role of Peccei-Quinn (PQ) symmetry \cite{pq}. 
If the spontaneous breaking of this $U(1)_{PQ}$ symmetry occurs, 
a pseudo Nambu-Goldstone boson associated to this breaking 
called axion appears to solve the strong $CP$ problem \cite{axion}.
If the axion decay constant $f_a$ is large enough such as 
$10^9~{\rm GeV}<f_a<10^{12}~{\rm GeV}$ due to a vacuum expectation 
value (VEV) of the singlet scalar, 
the axion mass is very small and its coupling is extremely weak
so as not to cause any contradiction with experiments and 
astrophysical observations \cite{fa}.

On the other hand, the $U(1)_{PQ}$ breaking is known to cause 
$N$ degenerate minima for the axion potential due to 
the QCD anomaly depending on both the field contents and the PQ 
charge assignment for them. 
As a result, the model is generally annoyed by the dangerous production of 
topologically stable domain walls \cite{dw}.  
It can be escapable only for $N=1$ unless one consider the domain wall 
free universe brought about by inflation.
If a certain subgroup of $U(1)_{\rm PQ}$ remains as a discrete symmetry
broken only by the QCD anomaly in a model with $N=1$, 
it could present an interesting scenario in relation to 
the DM physics at the low energy regions.\footnote{The similar idea has 
been discussed in several articles, recently \cite{pqscot}. 
However, the present model is different from them.} 

In this paper, we consider such a possibility in an extension of 
the KSVZ model, in which an inert doublet scalar and three 
right-handed neutrinos are added. 
The low energy effective model obtained from it 
after the breakdown of the $U(1)_{PQ}$ symmetry is reduced to the 
original scotogenic neutrino mass model with an effective $Z_2$ 
symmetry \cite{ma}. 
This $Z_2$ symmetry could guarantee the stability of a lightest 
neutral component of the inert doublet scalar to give a DM candidate.
The neutrino masses are generated through a one-loop effect
as a result of the $U(1)_{PQ}$ breaking.
The relevant diagram is caused by both right-handed neutrinos
and a nonrenormalizable interaction between the inert doublet scalar 
and the ordinary Higgs doublet.  
The model might be recognized as a well motivated simple framework 
at high energy regions for the original scotogenic model.

The remaining parts are organized as follows. In the next section,
we introduce a model by fixing charge assignment of $U(1)_{PQ}$ to
the field contents. We discuss basic features of the model such as 
remnant effective symmetry, scalar mass spectrum, 
vacuum stability and so on. 
In section 3, phenomenological features such as neutrino mass generation, 
leptogenesis and DM abundance in this model are discussed. 
The consistency of the scenario is also studied from a viewpoint of 
the vacuum stability and a cut-off scale of the model. 
We summarize the paper in section 4.

\section{An extension of the KSVZ model}
The KSVZ model is constructed by introducing a singlet complex scalar 
$S$ and a vector-like colored fermions $(D_L, D_R)$ to the SM \cite{ksvz}.
We assume $D_{L,R}$ as triplets of the color $SU(3)$.
Although they are $SU(2)_L$ singlets,
they could have a suitable weak hypercharge $Y$, in general. 
This point is crucial for phenomenological consistency of the model 
as discussed below.
The model has a global $U(1)_{PQ}$ symmetry and its charge is
assigned to $S$ and $D_{L,R}$, but it is not assigned to the SM contents.
We assume the existence of a gauge invariant Yukawa coupling 
$y_DS\bar D_LD_R$ so that the PQ mechanism could work to solve
the strong $CP$ problem.
This requires that the PQ charge $X$ of these new ingredients should 
satisfy $X_S=X_{D_L}-X_{D_R}$.  On the other hand, this symmetry should 
be chiral to have the QCD anomaly and $X_{D_L}\not=X_{D_R}$ is satisfied.
Thus, this $U(1)_{PQ}$ is spontaneously broken through the VEV of $S$. 

The $U(1)_{PQ}$ transformation $D_{L,R}\rightarrow e^{iX_{D_{L,R}}\alpha}D_{L,R}$ 
for the colored fermions $D_{L,R}$ shifts the QCD $\theta$ parameter through 
the anomaly as \cite{strongcp,dw}
\begin{equation}
\theta_{\rm QCD}\rightarrow \theta_{\rm QCD}-\frac{1}{2}(X_{D_R}-X_{D_L})\alpha.
\end{equation}
Since $\theta_{\rm QCD}$ has a period $2\pi$,  
the model is invariant for $\alpha=\frac{2\pi k}{N}$ where 
$N\equiv\frac{1}{2}|X_{D_R}-X_{D_L}|$ is an integer and $k=0,1,\cdots,N-1$.
This means that the model could have a discrete symmetry $Z_N$
after taking account of the QCD anomaly.\footnote{The axion decay constant 
$f_a$ is related with the PQ symmetry breaking scale $\langle S\rangle$ 
as $Nf_a=\langle S\rangle$ by using this $N$.}
If we assign the $U(1)_{PQ}$ charge $S$ as $X_S=2$,
the model has $N=1$ and no degenerate minima in the axion potential.
Thus, the model has no domain wall problem as is well known.\footnote{
Although the model has domain walls 
bounded by the string caused from the spontaneous $U(1)_{PQ}$ breaking,
it is not topologically stable and then it can shrink and decay. As a result,
no cosmological difficulty appears \cite{stdw}.}
Here, we note that an effective $Z_2$ symmetry could remain after 
the symmetry breaking due to $\langle S\rangle\not=0$ although it is
violated by the QCD anomaly. 
Since the SM contents are supposed to have no PQ charge,
it could play an important role in the leptonic sector of the model
to guarantee the stability of the lightest $Z_2$ odd field in that sector, 
which could be DM.
 
If both $D_L$ and $D_R$ cannot couple with quarks, which occurs in 
case $Y(D_{L,R})=0$ for example, they are stable and then
its relic abundance has to be smaller than the DM abundance \cite{lmn}.
Even if its relic abundance satisfies such a condition, the existence of 
the fractionally charged $D$ hadrons is generally forbidden by the present 
bound obtained from the search of fractionally charged states.
On the other hand, if we assign $Y=-\frac{1}{3}$ or $\frac{2}{3}$ to $D_{L,R}$, 
all the $D$ hadrons can have integer charge. In that case, the $D$ relic 
abundance will restrict the $D$ mass into a narrow range such as 
$m_D~{^>_\sim}~1$ TeV \cite{lmn}.
Moreover, they are allowed to couple with quarks through 
a renormalizable Yukawa interaction as long as their PQ charge is zero. 
For example, using the left handed 
quark doublet $q_L$ and the Higgs doublet $\phi$ or 
$\tilde\phi(\equiv i\tau_2\phi^\ast)$, the coupling 
$\tilde\phi\bar q_LD_R$ is allowed for $D_R$ with $X=0$ and $Y=-\frac{1}{3}$
and also $\phi\bar q_L D_R$ for $D_R$ with $X=0$ and $Y=\frac{2}{3}$.
In these cases, $D_R$ decays to the SM fields through these couplings. 
$D_L$ can also decay via the mass mixing with $D_R$ induced by 
the coupling $y_DS\bar D_LD_R$ through $\langle S\rangle\not=0$. 
As a result, the mass $m_D$ has no constraint other 
than the bound obtained through the accelerator experiments.
Anyway, in the model where the PQ charge is assigned as discussed above, 
the strong $CP$ problem could be solved without inducing 
any cosmological and astrophysical difficulty,
as long as the symmetry breaking scale satisfies 
$10^9~{\rm GeV}<\langle S\rangle<10^{12}~{\rm GeV}$.

Now, we consider a modification of this model by introducing an inert 
doublet scalar $\eta$ and three right-handed neutrinos $N_i$.
The PQ charge assignment of the fields contained in the model 
is shown in Table 1.
Invariant terms under the assumed symmetry for the Yukawa couplings 
and the scalar potential of the relevant fields are summarized as 
\begin{eqnarray}
-{\cal L}_y&=&y_DS\bar D_LD_R + h_D\bar q_L\tilde\phi D_R + y_iS\bar N^c_iN_i
+h_{\alpha i}\bar \ell_\alpha\eta N_i +{\rm h.c.}, \nonumber \\
V&=&m_S^2S^\dagger S+\kappa_1(S^\dagger S)^2+\kappa_2(S^\dagger S)(\phi^\dagger\phi)
+\kappa_3(S^\dagger S)(\eta^\dagger\eta) \nonumber \\
&+&m_\eta^2\eta^\dagger\eta +m_\phi^2\phi^\dagger\phi
+\lambda_1(\phi^\dagger\phi)^2
+\lambda_2(\eta^\dagger\eta)^2 
+\lambda_3(\phi^\dagger\phi)(\eta^\dagger\eta) 
+\lambda_4(\phi^\dagger\eta)(\eta^\dagger\phi) \nonumber \\ 
&+&\frac{\lambda_5}{2}\left[\frac{S}{M_\ast}(\eta^\dagger\phi)^2
+{\rm h.c.}\right],
\label{smodel}
\end{eqnarray}
where $\lambda_5$ is taken to be real and $M_\ast$ is 
a cut-off scale of the model. 
The quark generation index is abbreviated in the Yukawa coupling 
$h_D$.
We find that $V$ given in eq.~(\ref{smodel}) is the most general scalar 
potential up to the dimension 5.

\begin{figure}[t]
\begin{center} 
\begin{tabular}{c||cccccccccc}\hline
 &$D_L$ &$D_R$ & $S$ & $\eta$ & $N_i$ \\ \hline 
$Y$ & $-\frac{1}{3}$ & $-\frac{1}{3}$ & 0 & $-\frac{1}{2}$ & 0 \\
$X$ & 2 & 0 & 2& 1& $-1$  \\ 
$Z_2$ & $+$ & $+$ & $+$ & $-$ & $-$ \\ \hline
\end{tabular}
\end{center}

{\footnotesize {\bf Table~1}~~ The hypercharge $Y$ and the $U(1)_{\rm PQ}$ 
charge $X$ of new fields in the model. The SM contents are assumed to have 
no PQ charge. Parity for the effective symmetry $Z_2$ which remains 
after the $U(1)_{PQ}$ breaking is also listed.}  
\end{figure}

After the symmetry breaking due to $\langle S\rangle\not=0$, $D_{L,R}$, $N_i$
and $S$ are found to get masses such as 
$m_D=y_D\langle S\rangle$, $M_i=y_i\langle S\rangle$
and $M_S^2=4\kappa_1\langle S\rangle^2$, respectively.
Since $D_{L,R}$ can decay to the SM fields through the second term in 
${\cal L}_y$ as discussed above, there is no thermal relic 
of $D_{L,R}$ in the present Universe.
The effective model at the scale below $M_S$ could be obtained 
by integrating out $S$ \cite{stab}.
This can be done by using the equation of motion for $S$.
As its result, we obtain the corresponding
effective model whose scalar potential of the light 
scalars can be written as
\begin{eqnarray}
V_{\rm eff}&=&\tilde m_\phi^2(\phi^\dagger\phi)+\tilde m_\eta^2(\eta^\dagger\eta)
+\tilde\lambda_1(\phi^\dagger\phi)^2
+\tilde\lambda_2(\eta^\dagger\eta)^2
+\tilde\lambda_3(\phi^\dagger\phi)(\eta^\dagger\eta) 
+\lambda_4(\phi^\dagger\eta)(\eta^\dagger\phi) \nonumber \\
&+&\frac{\tilde\lambda_5}{2}\left[(\phi^\dagger\eta)^2 +{\rm h.c.}\right],
\label{effpot}
\end{eqnarray}
where we use the shifted parameters which are defined as 
\begin{eqnarray}
&&\tilde\lambda_1=\lambda_1-\frac{\kappa_2^2}{4\kappa_1}, \qquad
\tilde\lambda_2=\lambda_2-\frac{\kappa_3^2}{4\kappa_1}, \qquad
\tilde\lambda_3=\lambda_3-\frac{\kappa_2\kappa_3}{2\kappa_1}, \nonumber\\
&&\tilde\lambda_5=\lambda_5\frac{\langle S\rangle}{M_\ast}, \qquad
\tilde m_\phi^2=m_\phi^2+\kappa_2\langle S\rangle^2, \qquad 
\tilde m_\eta^2=m_\eta^2+\kappa_3\langle S\rangle^2.
\label{gcoupl}
\end{eqnarray}
We note that the model contains the neutrino 
Yukawa couplings between heavy right-handed 
neutrinos and the inert doublet scalar as shown in the above ${\cal L}_y$.

Vacuum stability condition for the scalar potential $V_{\rm eff}$ 
in eq.~(\ref{effpot}) is known to be given as \cite{pstab}
\begin{equation}
\tilde\lambda_1>0, \qquad \tilde\lambda_2>0, \qquad
\tilde\lambda_3>-2\sqrt{\tilde\lambda_1\tilde\lambda_2},  \qquad
\tilde\lambda_3+\lambda_4-|\tilde\lambda_5|>
-2\sqrt{\tilde\lambda_1\tilde\lambda_2},
\label{instab}
\end{equation}
and these should be satisfied at the energy region $\mu<M_S$.
On the other hand, at $M_S<\mu<M_\ast$, both the same conditions 
for $\lambda_{1,2,3}$ as eq.~(\ref{instab}) except for the last one 
and new conditions    
\begin{equation}
\kappa_1>0, \qquad
\kappa_2>-2\sqrt{\lambda_1\kappa_1}, \qquad
\kappa_3>-2\sqrt{\lambda_2\kappa_1},
\label{stability2}
\end{equation}
should be satisfied.
The couplings in both regions should be connected through eq.~(\ref{gcoupl}).
We can examine whether these conditions could be satisfied or not
by using one-loop renormalization group equations (RGEs).
This is the subject studied later. 
 
This effective model obtained after the spontaneous 
breaking of $U(1)_{PQ}$ is just the original scotogenic model 
 \cite{ma}.\footnote{In the case of $Y(D_{L,R})\not=0$,
$U(1)_{PQ}$ and then its subgroup $Z_2$ could be broken by the 
electroweak anomaly also.
However, since this breaking does not induce the decay of the
lightest $Z_2$ odd field, this $Z_2$ can be considered to be a good
symmetry in the effective model.} 
This model connects the neutrino mass generation with the DM existence. 
It has been extensively studied from various phenomenological 
view points \cite{radnm,tribi,stabf1,ks,infl}.
In the present case, the right-handed neutrinos do not have their masses
in a TeV region but they are considered to be much heavier.
The coupling $\tilde\lambda_5$ which is crucial for 
the one-loop neutrino mass generation is derived from a nonrenormalizable 
term as a result of the PQ symmetry breaking. 
The model contains the inert doublet scalar 
$\eta$ which has odd parity of the remnant effective $Z_2$.  
It has charged components $\eta^\pm$ and two neutral components $\eta_{R,I}$.
Their mass eigenvalues can be expressed as
\begin{equation}
M_{\eta^\pm}^2=\tilde m_\eta^2 +\tilde\lambda_3\langle\phi\rangle^2, \qquad
M_{\eta_{R,I}}^2=\tilde m_\eta^2+\left(\tilde\lambda_3+\lambda_4
\pm\tilde\lambda_5\right)\langle\phi\rangle^2.
\label{mscalar}
\end{equation}
We suppose $\tilde m_\eta=O(1)$~TeV although it requires fine tuning 
because of $|\langle S\rangle|\gg |\langle\phi\rangle|$. 
As a result of the effective $Z_2$ symmetry, 
the lightest one among the components of $\eta$ is stable 
to be a DM candidate if it is neutral.
If it is supposed to be $\eta_R$, we find that this requires 
$\lambda_4<0$ and $\tilde\lambda_5<0$
as long as $|\tilde\lambda_5|\ll |\lambda_4|$ is satisfied. 
On the other hand, since $\tilde m_\eta^2\gg \langle\phi\rangle^2$ is satisfied 
in eq.~(\ref{mscalar}), the mass eigenvalues of the components $\eta$ 
are found to be degenerate enough so that the coannihilation processes 
among them are expected to be effective. 
This observation suggests that the abundance of $\eta_R$ could be suitably 
suppressed and then it could be a good DM candidate as the ordinary 
inert doublet model \cite{inert1,inert2}. The charged states with 
the mass of $O(1)$~TeV are also expected to be detected 
in the accelerator experiments. 

\section{Phenomenological features} 
\subsection{Neutrino mass, leptogenesis and DM relic abundance}
In this model, neutrino masses are forbidden at tree-level.
However, since both the right-handed neutrino masses and the mass 
difference between $\eta_R$ and $\eta_I$ are induced after the $U(1)_{PQ}$ 
breaking, the small neutrino masses can be generated radiatively through 
one-loop diagrams in the same way as the original scotogenic model.
Since $M_{\eta_{R,I}}^2\gg |M_{\eta_R}^2-M_{\eta_I}^2|$ is satisfied, 
the neutrino mass formula can be approximately written as
\begin{equation}
{\cal M}_{\alpha\beta}=\sum_i h_{\alpha i}h_{\beta i}\Lambda_i, \qquad
\Lambda_i\simeq \frac{\tilde\lambda_5\langle\phi\rangle^2}{8\pi^2M_i}
\ln\frac{M_i^2}{\bar M_\eta^2},
\label{lnmass}
\end{equation} 
where 
$\bar M_\eta^2= \tilde m_\eta^2
+\left(\tilde\lambda_3+\lambda_4\right)\langle\phi\rangle^2$. 
In order to take account of the constraints from the neutrino 
oscillation data in the analysis, we may fix the flavor structure of
neutrino Yukawa couplings $h_{\alpha i}$ at the one which induces the 
tri-bimaximal mixing \cite{tribi}\footnote{Although a certain modification is 
required to reproduce the observed mixing in the lepton sector, 
this simplified example could give a rather good approximation 
for the present purpose as found from \cite{ks}.} 
\begin{equation}
h_{ej}=0, \quad h_{\mu j}=h_{\tau j}\equiv h_j \quad (j=1,2); \qquad 
h_{e3}=h_{\mu 3}=-h_{\tau 3}\equiv h_3, 
\label{flavor}
\end{equation}
where the charged lepton mass matrix is assumed to be diagonal.
In that case, the mass eigenvalues are estimated as 
\begin{eqnarray}
&&m_1=0, \qquad m_2= 3|h_3|^2\Lambda_3, \nonumber \\
&&m_3=2\left[|h_1|^4\Lambda_1^2+|h_2|^4\Lambda_2^2+
2|h_1|^2|h_2|^2\Lambda_1\Lambda_2\cos 2(\theta_1-\theta_2)
\right]^{1/2}, 
\label{nmass}
\end{eqnarray}
where $\theta_j={\rm arg}(h_j)$.

As is known generally and found also from this mass formula, 
neutrino masses could be determined only by two right-handed neutrinos.
It means that the mass and neutrino Yukawa couplings of a remaining 
right-handed neutrino could be free from the neutrino oscillation data
as long as its contribution to the neutrino mass is negligible.
In eq.~(\ref{nmass}), such a situation can be realized for 
$|h_1|^2\Lambda_1 \ll |h_2|^2\Lambda_2$.
This is good for the thermal leptogenesis \cite{leptg} since a 
sufficiently small
neutrino Yukawa coupling $h_1$ makes the out-of-equilibrium decay of the
right-handed neutrino $N_1$ possible.\footnote{If we consider the TeV 
scale right-handed neutrinos, leptogenesis requires fine degeneracy 
among the right-handed neutrinos for the resonance \cite{resonance}. 
We need not consider such a possibility in the present case.}
We find that the squared mass differences required by the 
neutrino oscillation data could be explained if we fix the parameters 
relevant to the neutrino masses, for example, as
\begin{eqnarray}
&&M_1= 10^8~{\rm GeV}, \qquad M_2=4\times 10^8~{\rm GeV}, 
\qquad M_3= 10^9{\rm GeV},  \nonumber \\
&&|h_1|= 10^{-4.5},  
\qquad |h_2|\simeq 7.2\times 10^{-4}\tilde\lambda_5^{-0.5}, \qquad 
|h_3|\simeq 3.1\times 10^{-4}\tilde\lambda_5^{-0.5}, 
\label{yukawa}
\end{eqnarray}
for $\tilde m_\eta=1$~TeV.
Using these values, we can estimate the expected baryon number asymmetry
through the out-of-equilibrium decay of the thermal $N_1$ by solving the
Boltzmann equation as done in \cite{ks}. 
The numerical analysis shows that the required 
baryon number asymmetry could be generated for $M_1~{^>_\sim}~10^8$~GeV, 
which is somewhat smaller than the Davidson-Ibarra bound \cite{di}
in the ordinary thermal leptogenesis.
In case of the parameter set given in (\ref{yukawa}), we find 
$Y_B\left(\equiv\frac{n_B}{s}\right)
=4.0\times 10^{-10}$ if we assume $\tilde\lambda_5=2.5\times 10^{-3}$
and a maximal $CP$ phase in the $CP$ violation parameter $\varepsilon$. 
In Fig.~1, we plot $Y_B$ as a function of $\tilde\lambda_5$.
Its feature can be easily understood by taking account of eq.~(\ref{yukawa}).
If $\tilde\lambda_5$ takes larger values, the neutrino Yukawa couplings
become smaller to make the $CP$ violation $\varepsilon$ in the
$N_1$ decay smaller but also the washout of the generated lepton number 
asymmetry smaller. On the other hand, if $\tilde\lambda_5$ takes smaller 
values, the neutrino Yukawa couplings become larger to induce the 
reverse effects. 
This makes the required baryon number asymmetry be generated 
only for the $\tilde\lambda_5$ in the limited 
regions as found in this figure.

\input epsf
\begin{figure}[t]
\begin{center}
\epsfxsize=7.5cm
\leavevmode
\epsfbox{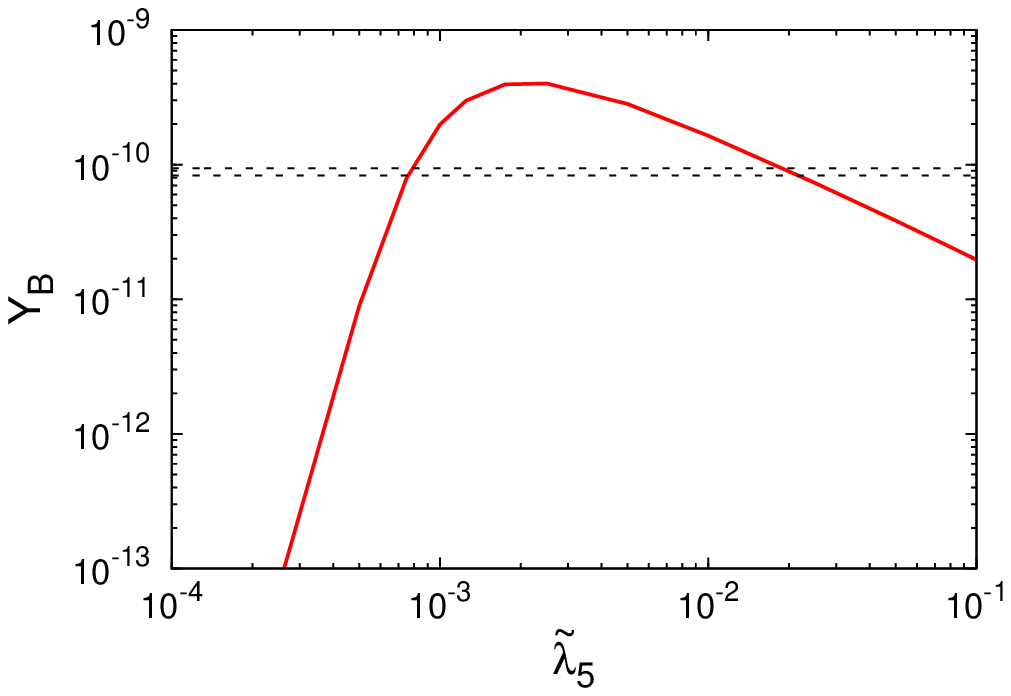}
\end{center}
\vspace*{-3mm}

{\footnotesize {\bf Fig.~1}~~Baryon number asymmetry 
$Y_B$ generated through 
the out-of-equilibrium decay of $N_1$. $Y_B$ is plotted as a function of 
$\tilde\lambda_5$ for the parameter set shown in eq.~(\ref{yukawa}), which 
can explain the neutrino mass differences required 
by the neutrino oscillation data. Horizontal dotted lines show 
the required value for $Y_B$. }
\end{figure}

The relic abundance of $\eta_R$ is tuned to the observed value 
if the couplings $\tilde\lambda_3$ and $\lambda_4$ 
take suitable values.
In fact, since $\tilde m_\eta$ is assumed to be of
$O(1)$~TeV in this scenario, the mass of each component of $\eta$ could 
be degenerate enough for wide range values of $\tilde\lambda_3$ and 
$\lambda_4$ as remarked at eq.~(\ref{mscalar}). 
This makes the coannihilation among them effective enough to reduce 
the $\eta_R$ abundance \cite{ks}.
We search the region of $\tilde\lambda_3$ and $\lambda_4$,
which realizes the required DM abundance as the $\eta_R$ relic abundance 
by taking the values of $\tilde m_\eta$ and $\tilde\lambda_5$ 
as the ones given below eq.~(\ref{yukawa}).
They are suitable for the explanation of the neutrino oscillation data and
the cosmological baryon number asymmetry.
In the estimation of the DM relic abundance, we follow the procedure 
given in \cite{gs} where the coannihilation effects are taken into account.

We present a brief review of the procedure adopted here.
The $\eta_R$ relic abundance is estimated as
\begin{equation}
\Omega h^2\simeq \frac{1.07\times 10^9~{\rm GeV}^{-1}}{J(x_F)g_\ast^{1/2}m_{\rm pl}},
\end{equation}
where $g_\ast$ is the relativistic degrees of freedom. 
The freeze-out temperature $T_F(\equiv\frac{M_{\eta_R}}{x_F})$ of $\eta_R$
and $J(x_F)$ are defined as
\begin{equation}
x_F=\ln\frac{0.038m_{\rm pl}g_{\rm eff}M_{\eta_R}\langle\sigma_{\rm eff}v\rangle}
{(g_\ast x_F)^{1/2}}, \qquad
J(x_F)=\int^\infty_{x_F}\frac{\langle\sigma_{\rm eff}v\rangle}{x^2}dx.
\end{equation}
In these formulas, the effective annihilation cross section 
$\langle\sigma_{\rm eff}v\rangle$
and the effective degrees of freedom $g_{\rm eff}$ are expressed 
as\footnote{In this part, we label $(\eta_R,\eta_I,\eta^+,\eta^-)$ 
as $(\eta_1,\eta_2,\eta_3,\eta_4)$.}
\begin{equation}
\langle\sigma_{\rm eff}v\rangle=\frac{1}{g_{\rm eff}^2}\sum^4_{i,j=1}
\langle\sigma_{ij}v\rangle\frac{n^{\rm eq}_i}{n^{\rm eq}_1}
\frac{n^{\rm eq}_j}{n^{\rm eq}_1}, \qquad
g_{\rm eff}=\sum_{i=1}^4\frac{n^{\rm eq}_i}{n^{\rm eq}_1}, 
\end{equation}
where $\langle\sigma_{ij}v\rangle$ is the thermally averaged 
(co)annihilation cross section and $n^{\rm eq}_i$ is the thermal 
equilibrium number density of $\eta_i$.
If the former is expanded by the thermally averaged relative velocity 
$\langle v^2\rangle$ as 
$\langle\sigma_{ij}v\rangle = a_{ij}+b_{ij}\langle v^2\rangle$,
it could be approximated only by 
$a_{ij}$ since $\langle v^2\rangle\ll1$ is satisfied for the cold DM.
Final states of the relevant (co)annihilation are
composed only of the SM contents. 
The corresponding $a_{\rm eff}$ can be approximately 
calculated as \cite{ks,inert2}
\begin{eqnarray}
&&a_{\rm eff}=\frac{(1+2c_w^4)g^4}{128\pi
 c_w^4M_{\eta_1}^2}\left(N_{11}+N_{22}+ 2N_{34}\right)
+\frac{s_w^2g^4}{32\pi c_w^2M_{\eta_1}^2}
\left(N_{13}+N_{14}+N_{23}+N_{24}\right) \nonumber \\
&&+\frac{1}{64\pi M_{\eta_1}^2}\left[
\left(\tilde\lambda_+^2+\tilde\lambda_-^2
+2\tilde\lambda_3^2\right)(N_{11}+N_{22}) 
+(\tilde\lambda_+-\tilde\lambda_-)^2(N_{33}+N_{44}+ N_{12})\right. \nonumber \\
&&+\left\{(\tilde\lambda_+-\tilde\lambda_3)^2
+(\tilde\lambda_--\tilde\lambda_3)^2\right\}(N_{13}+N_{14}+N_{23}+N_{24})
\nonumber \\
&&+\left.\left\{(\tilde\lambda_+ +\tilde\lambda_-)^2
+4\tilde\lambda_3^2\right\}N_{34}\right], 
\label{cross}
\end{eqnarray}
where $\tilde\lambda_\pm=\tilde\lambda_3+\lambda_4\pm\tilde\lambda_5$ and 
$N_{ij}$ is defined by using $M_{\eta_i}$ given in eq.~(\ref{mscalar}) as
\begin{equation}
N_{ij}\equiv\frac{1}{g_{\rm eff}^2}
\frac{n_i^{\rm eq}}{n_1^{\rm eq}}\frac{n_j^{\rm eq}}
{n_1^{\rm eq}}
=\frac{1}{g_{\rm eff}^2}
\left(\frac{M_{\eta_i}M_{\eta_j}}{M_{\eta_1}^2}\right)^{3/2}
\exp\left[-\frac{M_{\eta_i}+M_{\eta_j}-2M_{\eta_1}}{T}\right].
\label{eqfactor}
\end{equation}

\begin{figure}[t]
\begin{center}
\epsfxsize=7.5cm
\leavevmode
\epsfbox{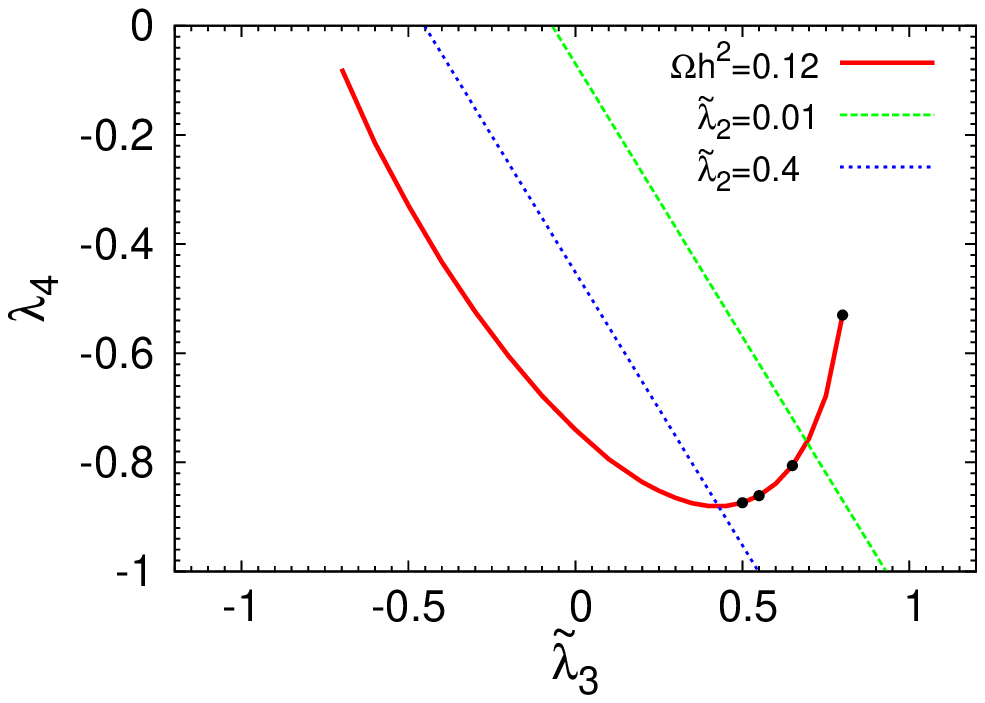}
\end{center}
\vspace*{-3mm}

{\footnotesize {\bf Fig.~2}~~
Points plotted by a red solid line in the 
$(\tilde\lambda_3,~\lambda_4)$ plane can realize
the required DM relic abundance $\Omega h^2=0.12$ as the relic
$\eta_R$ abundance. The last condition in eq.(\ref{instab})
is satisfied at a region above the straight line which represents
$\tilde\lambda_3+\lambda_4=|\tilde\lambda_5|-
2\sqrt{\tilde\lambda_1\tilde\lambda_2}$ for a fixed $\tilde\lambda_2$.}
\end{figure}

We use this procedure to find the points in the 
$(\tilde\lambda_3,~\lambda_4)$ plane, where the required DM 
abundance $\Omega_{\rm DM}h^2=0.12$ is realized by $\eta_R$.
In Fig.~2, we plot such points by a red solid line 
for $\tilde m_\eta=1$~TeV and $\tilde\lambda_5=2.5\times 10^{-3}$ 
which are used in the previous part.
In this figure, we take account of the condition $\lambda_4<0$ 
which has been already discussed in relation to eq.~(\ref{mscalar}).
Moreover, if we use the Higgs mass formula  
$m_{h^0}^2=4\tilde\lambda_1\langle\phi\rangle^2$, 
we find $\tilde\lambda_1\simeq 0.13$ for $m_{h^0}=125$~GeV
and then the last condition in eq.~(\ref{instab}) can be also 
plotted for a fixed $\tilde\lambda_2$ in the same plane.\footnote{We note that
the second condition in eq.~(\ref{instab}) is automatically satisfied if 
the last one is fulfilled.}
An allowed points are contained in the region above a straight line
which is fixed by an assumed value of $\tilde\lambda_2$.
We give two examples here. 
Although the DM abundance can be satisfied for the negative value of
$\tilde\lambda_3$, we find that such cases contradict with the 
vacuum stability condition for $\tilde\lambda_3$ given in eq.~(\ref{instab}).
The figure shows that $\tilde\lambda_3$ and/or $|\lambda_4|$ are 
required to take rather large values for realization of the DM abundance. 
This suggests that the RG evolution of the scalar quartic couplings 
$\tilde\lambda_i$ could be largely affected if they are used as 
initial values at the weak scale.
In that case, vacuum stability and perturbativity of the model could 
give constraints on the model. 
In the next part, we focus our study on this point. 

Before proceeding to this subject, we comment on the contribution 
of the axion to the DM abundance and also
a possible violation of $U(1)_{PQ}$ by the quantum gravity effect.
In this model, the axion could also contribute to the DM abundance 
through the misalignment mechanism. 
If the initial misalignment of the axion 
is written as $\langle\theta_i\rangle$,
the axion contribution to the present energy density is 
estimated as \cite{strongcp}
\begin{equation} 
\Omega_ah^2=2\times 10^4\left(\frac{\langle S\rangle}
{10^{16}~{\rm GeV}}\right)^{7/6}\langle\theta_i^2\rangle.
\end{equation}
The axion contribution to the DM abundance crucially depends on 
the scale of $\langle S\rangle$ and $\langle\theta_i\rangle$. 
This estimation shows that it could be too small to give the required value 
$\Omega_{\rm DM} h^2=0.12$ for $\langle S\rangle< 10^{11}$~GeV
even if we assume $\langle\theta_i\rangle=O(1)$.\footnote{The estimation of 
the relic axion abundance has to take account of the contribution from 
the decay of string and domain walls. 
Depending on it, the upper bound on the PQ breaking scale
seems to be somewhat ambiguous. 
While one group finds that the axion production is 
more efficient than the misalignment case \cite{axion1}, the other group finds
that it is less efficient than the misalignment case \cite{axion2}.}
Thus, the axion contribution to the DM abundance is sub-dominant 
or negligible for $\langle S\rangle< 10^{11}$~GeV.
In this region of $\langle S\rangle$, 
the result obtained for $(\tilde\lambda_3,\lambda_4)$ through 
the above study can be still applicable even if the axion contribution 
to the DM abundance is taken into account.
 
Although we assume that $U(1)_{PQ}$ is exact in this study, 
continuous global symmetry is suggested to be violated by the quantum gravity. 
This possible effect on the PQ mechanism has been 
studied \cite{gravity}.
If the $U(1)_{\rm PQ}$ symmetry is violated by the gravity induced effective 
interaction which is suppressed by the Planck scale such as
\begin{equation}
\frac{|S|^{n+3}}{M_{\rm pl}^n}\left(gS+g^\ast S\right),
\end{equation}
it has been shown that $n\ge 6$ should be satisfied for the PQ mechanism
to give a solution to the strong $CP$ problem in case that $|g|$ 
is of $O(1)$. If accidental appearance of global $U(1)$ happens due to 
some discrete or continuous gauge symmetry \cite{disgauge}, 
it might protect the PQ symmetry up to sufficiently higher order operators. 
The same breaking effect could also affect the axion CDM 
abundance \cite{gravity}.
If the contribution to the axion mass due to the quantum gravity 
is small compared to the one due to the QCD anomaly, 
$\langle S\rangle\simeq 10^{11}$~GeV is required for saturating 
$\Omega h^2=0.12$ by the axion contribution.
Even if its contribution to the axion mass is larger than the one 
from the QCD anomaly within the bound which is required so as not 
to disturb the PQ mechanism, $\langle S\rangle\simeq 10^{11}$~GeV is required
again for saturating $\Omega h^2=0.12$.   
Thus, $\eta_R$ could play a dominant role in the DM abundance as long as 
$\langle S\rangle$ is smaller than $10^{11}$~GeV.
 
The stability for $\eta_R$ could be also violated through the same effect.
The most effective processes for the $\eta_R$ decay are induced by 
nonrenormalizable Yukawa couplings such as
\begin{equation}
\frac{S^{n}}{M_{\rm pl}^{n}}\left(h_u\bar q_Lu_R\eta+h_d\bar q_Ld_R\tilde\eta
+h_e\bar \ell_Le_R\tilde\eta\right).
\label{decay}
\end{equation}
If the allowed dimension for these kind of operators is the same as
the one which guarantees the PQ mechanism to work, the lifetime of $\eta_R$
could be longer than the age of our universe in case $m_\eta=O(1)$~TeV
and $h_{u,d,e}=O(1)$ as long as we take $\langle S\rangle\simeq 10^{10}$~GeV.
If the lower dimension operators such as $n< 6$ are allowed, 
its lifetime cannot be long enough to be the DM at the present universe. 

\subsection{Consistency of the scenario with a cut-off scale of the model}
It is crucial to check what kind of values of the right-handed neutrino 
mass $M_i$ and $\tilde\lambda_5$
could be consistent with a value of $\langle S\rangle$ which is restricted 
by the axion physics. 
In this model, DM is identified with $\eta_R$ whose mass is of $O(1)$~TeV.  
In such a mass region, we find that its abundance is determined 
by the values of the scalar quartic couplings $\tilde\lambda_3$ and 
$\lambda_4$.
On the other hand, these couplings could affect the vacuum stability 
and also the perturbativity of the model through the radiative effects on the
scalar quartic couplings $\tilde\lambda_i$.
Here, we examine the consistency of the values of $\tilde\lambda_3$ and
$\lambda_4$ required to realize of the DM abundance 
with these issues.\footnote{The constraint due to the vacuum stability 
and the perturbativity is taken into account in the DM study of 
the inert doublet model on the basis of a different viewpoint 
from the present one \cite{inert1,inert2}.
The consistency between fermionic DM and the vacuum stability is also studied 
in the scotogenic model \cite{stabf1,stabf2}.}   
Since the breaking of the perturbativity is considered to be relevant to 
a scale for the applicability of the model, we could obtain an information for 
the cut-off scale $M_\ast$. It allows us to judge whether 
the required value for $\tilde\lambda_5$ by the neutrino masses 
and the leptogenesis could be induced through the VEV of $S$.  

The one-loop $\beta$-functions for the scalar quartic couplings in 
the effective model at energy regions below $M_S$ are   
given as follows \cite{rge}, 
\begin{eqnarray}
\beta_{\tilde\lambda_1}&=&24\tilde\lambda_1^2
+\tilde\lambda_3^2+(\tilde\lambda_3+\lambda_4)^2
+\tilde\lambda_5^2 \nonumber\\
&+&\frac{3}{8}\left(3g^4+g^{\prime 4}+2g^2g^{\prime 2}\right)
-3\tilde\lambda_1\left(3g^2+g^{\prime 2}-4h_t^2\right)-6h_t^4, \nonumber \\
\beta_{\tilde\lambda_2}&=&24\tilde\lambda_2^2+\tilde\lambda_3^2
+(\tilde\lambda_3+\lambda_4)^2+\tilde\lambda_5^2 \nonumber\\
&+&\frac{3}{8}\left(3g^4+g^{\prime 4}+2g^2g^{\prime 2}\right)
-3\tilde\lambda_2\left(3g^2+g^{\prime 2}\right), \nonumber \\
\beta_{\tilde\lambda_3}&=&2(\tilde\lambda_1+\tilde\lambda_2)
(6\tilde\lambda_3+2\lambda_4)
+4\tilde\lambda_3^2+2\lambda_4^2+2\tilde\lambda_5^2 \nonumber\\
&+&\frac{3}{4}\left(3g^4+g^{\prime 4}-2g^2g^{\prime 2}\right)
-3\tilde\lambda_3\left(3g^2+g^{\prime 2}-2h_t^2\right), \nonumber \\
\beta_{\lambda_4}&=&4(\tilde\lambda_1+\tilde\lambda_2)\lambda_4
+8\tilde\lambda_3\lambda_4+4\lambda_4^2
+8\tilde\lambda_5^2+3g^2g^{\prime 2}
-3\lambda_4\left(3g^2+g^{\prime 2}-2h_t^2\right), \nonumber\\
\beta_{\tilde\lambda_5}&=&4(\tilde\lambda_1+\tilde\lambda_2)\tilde\lambda_5
+8\tilde\lambda_3\tilde\lambda_5+12\lambda_4\tilde\lambda_5 
-3\tilde\lambda_5\left(3g^2+g^{\prime 2}-2h_t^2\right), 
\end{eqnarray}
where $\beta_\lambda$ is defined as 
$\beta_\lambda=16\pi^2\mu\frac{d\lambda}{d\mu}$. 
In these equations, we can expect that the positive contributions 
of $\tilde\lambda_3$ and $\lambda_4$ to the $\beta$-functions 
of $\tilde\lambda_{1,2}$ tend to save the model from violating the first 
two vacuum stability conditions in eq.~(\ref{instab}).
On the other hand, the same contributions of $\tilde\lambda_3$ and 
$\lambda_4$ could induce the breaking of the perturbativity of the model 
at a rather low energy scale since they could give large positive 
contributions to $\beta_{\tilde\lambda_1}$, $\beta_{\tilde\lambda_2}$ and 
$\beta_{\tilde\lambda_3}$.
Here, we identify a cut-off scale $M_\ast$ of the model with a scale 
where any of the perturbativity conditions $\lambda_i(M_\ast)<4\pi$ and
$\kappa_i(M_\ast)<4\pi$ is violated.\footnote{Since the Landau pole 
appearing scale is expected 
to be near to this $M_\ast$, it seems to be natural to identify $M_\ast$ 
with a cut-off scale of the model.}
In this case, $M_\ast >|\langle S\rangle|$ should be satisfied.
If $M_\ast$ is smaller than $\langle S\rangle$, the consistency of 
the scenario is lost. 

We analyze this issue by solving the above one-loop RGEs at $\mu<M_S$ and also
the ones at $\mu>M_S$, which are given in Appendix.
The quartic couplings $\tilde\lambda_i$ 
in the tree-level potential at the energy scale $\mu<M_S$ are connected 
with the ones $\lambda_i$ at $\mu>M_S$ through eq.~(\ref{gcoupl}).
Since the masses of the right-handed neutrinos $N_i$ are considered 
to be heavy in the present model, they decouple 
at the scale $\mu<M_i~^<_\sim~O(M_S)$ to be irrelevant to the RGEs there.
On the other hand, the mass of the colored fields $D_{L,R}$ can take any 
values larger than 1~TeV as discussed before, they can contribute to the 
RGEs at larger scales than their mass. 
In the present study, we assume that $D_{L,R}$ is light of $O(1)$~TeV 
but its Yukawa coupling $h_D$ with the ordinary quarks 
is small enough.\footnote{In the light $D$ case, study of the bound for 
this Yukawa coupling is an interesting subject related to the search of 
mixing with the ordinary quarks. However, it is beyond the scope of 
the present study and we do not discuss it here.}
Thus, they are considered to contribute substantially 
only to the $\beta$-functions of the gauge couplings.  
In this study, we take its hypercharge as $Y=-\frac{1}{3}$ as shown in Table 1.

The free parameters in the scalar potential of the effective model 
(\ref{effpot}) 
are $\tilde\lambda_1,~\tilde\lambda_2,~\tilde\lambda_3,~\lambda_4$ and 
$\tilde\lambda_5$ at $M_Z$ as long as 
we assume $\tilde m_\eta=1$~TeV.\footnote{Quartic couplings $\kappa_i$ 
for $S$ are fixed as $\kappa_1=\frac{M_S^2}{4\langle S\rangle^2}$ 
and $\kappa_{2,3}=0.1$ at $M_S$ in the present study.
As easily found from RGEs, larger values of $\kappa_{2,3}$ make $M_\ast$ 
smaller.}
Among them, we should fix $\tilde\lambda_5$ at a value used in the discussion 
of the neutrino mass and the leptogenesis.
Both $\tilde\lambda_3$ and $\lambda_4$ are fixed at values determined 
through the DM relic abundance as shown in Fig.~2. 
We also have $\tilde\lambda_1\simeq 0.13$ from the Higgs mass.
From this point of view, $\tilde\lambda_2$ is an only remaining parameter.
Thus, if we solve the RGEs varying the value of $\tilde\lambda_2$ 
for other fixed parameters, we can find $M_\ast$ checking 
the vacuum stability for each $\tilde\lambda_2$.

\begin{figure}[t]
\begin{center}
\epsfxsize=7.5cm
\leavevmode
\epsfbox{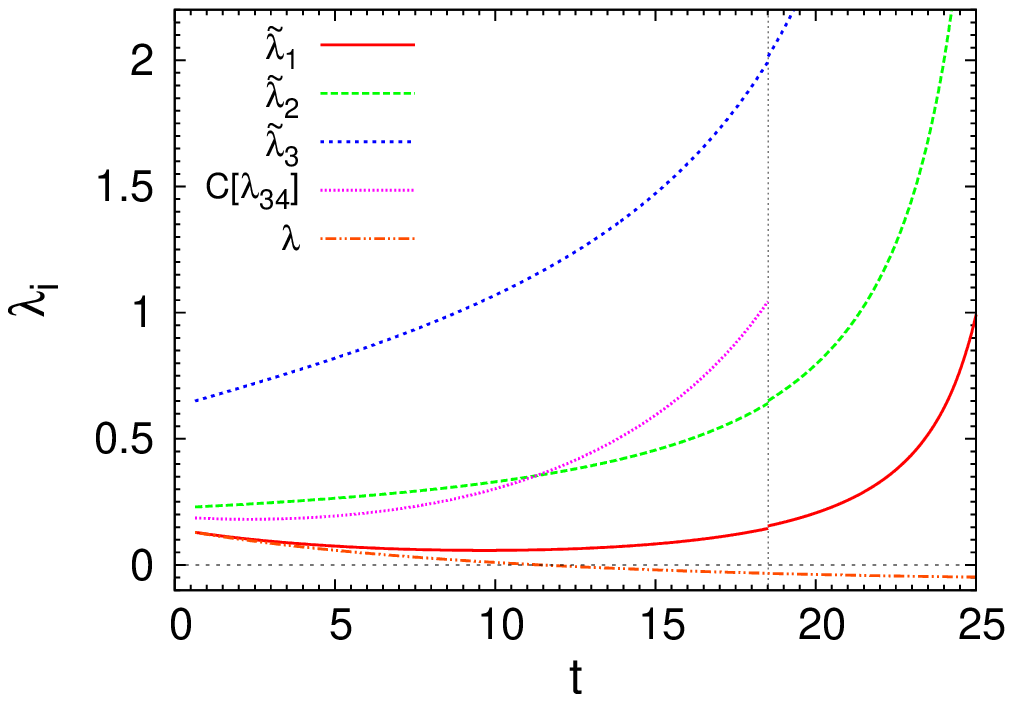}
\hspace*{5mm}
\epsfxsize=7.5cm
\leavevmode
\epsfbox{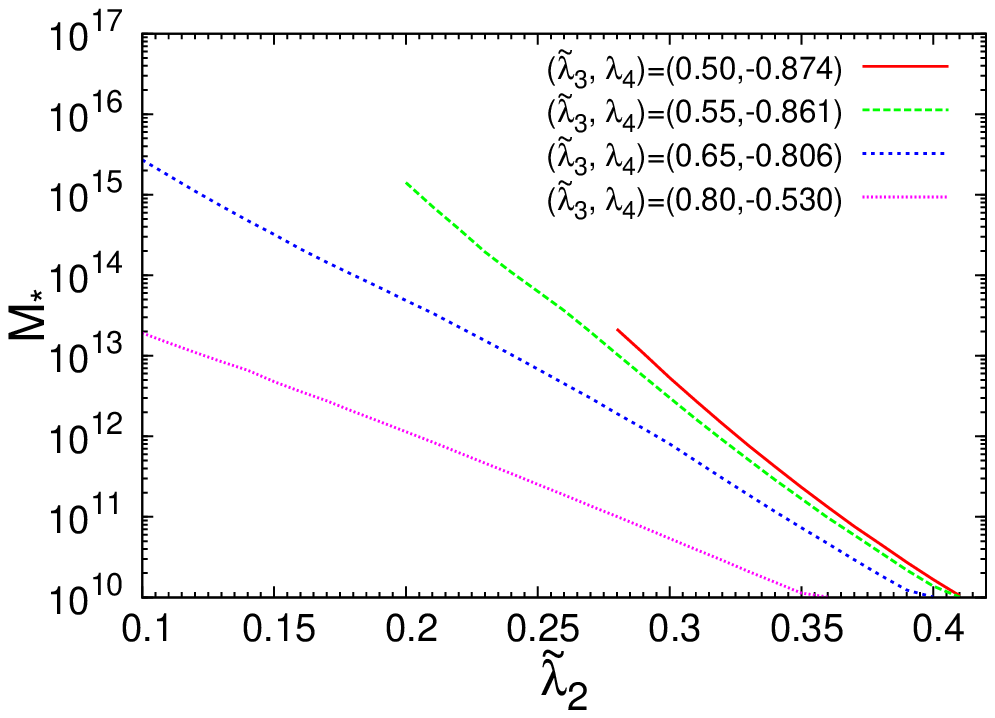}
\end{center}
\vspace*{-3mm}

{\footnotesize {\bf Fig.~3}~~
Left panel: running of the scalar quartic couplings for $t=\ln\frac{\mu}{M_Z}$.
$\tilde\lambda_2=0.23$, $\tilde\lambda_3=0.65$ and $\lambda_4=-0.806$ 
are used as the initial values at $\mu=M_Z$. 
A vertical line corresponds to
$t=\ln\left(\frac{M_S}{M_Z}\right)$. The running of the SM Higgs quartic 
coupling $\lambda$ is also plotted as a reference. 
Right panel: the cut-off scale $M_\ast$ as a function of $\tilde\lambda_2$
which is fixed as a value at $M_Z$ for four points marked 
by the black bulbs in Fig.~2 where $\Omega h^2=0.12$ is satisfied.}
\end{figure}

In the left panel of Fig.~3, as an example, we present the running 
of the scalar quartic couplings $\tilde\lambda_{1,2,3}$ 
for the  initial values $\tilde\lambda_2=0.23$, $\tilde\lambda_3=0.65$
and $\lambda_4=-0.806$ at $M_Z$ by assuming 
the $U(1)_{PQ}$ breaking scale as $\langle S\rangle=M_S=10^{10}$~GeV. 
In the same panel, we also plot the value of
$\tilde\lambda_3+\lambda_4-|\tilde\lambda_5|
+2\sqrt{\tilde\lambda_1\tilde\lambda_2}$ as $C[\lambda_{34}]$,
which corresponds to the last one in eq.~(\ref{instab}). 
In this example, we can see that the vacuum stability is kept until
the cut-off scale $M_\ast \simeq 1.54\times 10^{13}$~GeV.
These values of $\langle S\rangle$ and $M_\ast$ can naturally 
realize the assumed value for $\tilde\lambda_5$ through the relation given 
in eq.(\ref{gcoupl}) just by taking $\lambda_5$ as a value of $O(1)$. 
This feature can be verified for other allowed values
of $\tilde\lambda_3$ and $\lambda_4$.
Here, we note that the axion contribution to the DM abundance can 
be neglected for a value such as $\langle S\rangle<10^{11}$~GeV.
In the right panel of Fig.~3, we plot $M_\ast$ as a function 
of $\tilde\lambda_2$ for four sets of $(\tilde\lambda_3,~\lambda_4)$
which are shown by black bulbs in Fig.~2.
End points found in the two lines represent the value of $\tilde\lambda_2$
for which the vacuum stability is violated before reaching $M_\ast$. 
This figure shows that $\tilde\lambda_2$ which is 
restricted to a rather narrow region 
can make $M_\ast$ appropriate values in order to realize a required value 
of $\tilde\lambda_5$ for $\langle S\rangle<10^{11}$~GeV. 
This study suggests that the scenario could work well without
strict tuning of the relevant parameters.

As found from the above study, the simultaneous 
explanation of the neutrino masses and the DM abundance could be 
preserved in this extended model in the same way 
as in the original scotogenic model. 
We should stress that no other additional constraint from the DM physics 
and the neutrino physics is brought about by taking the present scenario. 
The cosmological baryon number asymmetry is expected to be explained through 
the out-of-equilibrium decay of the lightest right-handed neutrino.
The required right-handed neutrino mass could be smaller 
compared with the Davidson-Ibarra bound in the ordinary thermal 
leptogenesis \cite{di}. 
This is consistent with the result in \cite{ks} where the mass bound of 
the right-handed neutrino for the successful leptogenesis is shown to 
be relaxed in the radiative neutrino mass model 
in comparison with the ordinary seesaw model . 

Finally, we give brief comments on possible experimental signatures 
of the model. The present model might be examined through 
(i) the search of the $\eta_R$ DM and the charged scalars $\eta^\pm$ 
through the DM direct detection experiments and the accelerator experiments,
(ii) the search of the mixing of $D$ with the ordinary quarks 
although it could be observed only in the light $D$ case,
and (iii) the search of the axion whose coupling with photon
is characterized by $g_{a\gamma\gamma}=\frac{m_a}{\rm eV}
\frac{2.0}{10^{10}{\rm GeV}}(6 Y^2-1.92)$, where $Y$ 
is the hypercharge of $D$ \cite{lmn}.

\section{Summary}  
We have proposed an extension of the KSVZ invisible axion model
so as to include a DM candidate and explain the small neutrino masses.
An extra inert doublet scalar $\eta$ and 
three right-handed neutrinos $N_i$ are introduced as new ingredients.
After the $U(1)_{PQ}$ symmetry breaking, its subgroup $Z_2$ could remain 
as a remnant effective symmetry, which is violated through the QCD
anomaly but it can play the same role as the $Z_2$ in the scotogenic 
neutrino mass model. 
Since only the new ones $\eta$ and $N_i$ have its odd parity,
the model reduces to the scotogenic model which has $Z_2$ 
in the leptonic sector.
The neutrino masses are generated at one-loop level and the DM abundance
can be explained by the thermal relics of the neutral component of $\eta$.
The cosmological baryon number asymmetry could be generated through
the out-of-equilibrium decay of a right-handed neutrino in the same way 
as the ordinary thermal leptogenesis in the tree-level seesaw model.
However, the bound for the right-handed neutrino mass can be 
relaxed in this model.  
Since this simple extension can relate the strong $CP$ problem to the
origin of neutrino masses and DM, it may be a promising extension of 
both the KSVZ model and the scotogenic model.

\section*{Appendix}
The $\beta$-function for the scalar quartic couplings at $\mu>M_S$ are 
given as 
\begin{eqnarray}
\beta_{\lambda_1}&=&24\lambda_1^2
+\lambda_3^2+(\lambda_3+\lambda_4)^2 + \kappa_2^2
+\frac{3}{8}\left(3g^4+g^{\prime 4}+2g^2g^{\prime 2}\right) \nonumber \\
&-&3\lambda_1\left(3g^2+g^{\prime 2}-4h_t^2\right)-6h_t^4, \nonumber \\
\beta_{\lambda_2}&=&24\lambda_2^2+\lambda_3^2
+(\lambda_3+\lambda_4)^2 +\kappa_3^2 
+\frac{3}{8}\left(3g^4+g^{\prime 4}+2g^2g^{\prime 2}\right)
-3\lambda_2\left(3g^2+g^{\prime 2}\right)  \nonumber \\
&+&4\lambda_2\left[2(h_1^2+h_2^2)+3h_3^2\right]
-8(h_1^2+h_2^2)^2-18h_3^4, \nonumber \\
\beta_{\lambda_3}&=&2(\lambda_1+\lambda_2)
(6\lambda_3+2\lambda_4)
+4\lambda_3^2+2\lambda_4^2 +2\kappa_2\kappa_3  
+\frac{3}{4}\left(3g^4+g^{\prime 4}-2g^2g^{\prime 2}\right) \nonumber \\
&-&3\lambda_3\left(3g^2+g^{\prime 2}-2h_t^2\right)
+2\lambda_3\left[2(h_1^2+h_2^2)+3h_3^2\right], \nonumber \\
\beta_{\lambda_4}&=&4(\lambda_1+\lambda_2)\lambda_4
+8\lambda_3\lambda_4+4\lambda_4^2
+3g^2g^{\prime 2}-3\lambda_4\left(3g^2+g^{\prime 2}-2h_t^2\right) \nonumber \\
&+&2\lambda_4\left[2(h_1^2+h_2^2)+3h_3^2\right], \nonumber\\
\beta_{\kappa_1}&=& 20\kappa_1^2+2\kappa_2^2+2\kappa_3^2
+4\kappa_1\left(3y_D^2+\sum_iy_i^2\right)
-2\left(3y_D^4+\sum_iy_i^4\right), \nonumber \\
\beta_{\kappa_2}&=& 4\kappa_2^2+2\kappa_2(6\lambda_1+4\kappa_1)
+2\kappa_3(2\lambda_3+\lambda_4)+
2\kappa_2\left(3y_D^2+\sum_iy_i^2\right) \nonumber \\
&-&\frac{3}{2}\kappa_2(3g^2+g^{\prime 2}-4h_t^2), 
\nonumber \\
\beta_{\kappa_3}&=&4\kappa_3^2+2\kappa_3(6\lambda_2+4\kappa_1)
+2\kappa_2(2\lambda_3+\lambda_4)+
2\kappa_3\left(3y_D^2+\sum_iy_i^2\right) \nonumber \\
&-&\frac{3}{2}\kappa_3\left[3g^2+g^{\prime 2}-
\frac{4}{3}\left(2(h_1^2+h_2^2)+3h_3^2\right)\right], 
\end{eqnarray}
where eq.~(\ref{flavor}) is assumed for the flavor structure 
of neutrino Yukawa couplings.
The $\beta$-functions for the gauge couplings and the Yukawa couplings
for top, $D$ and neutrinos are given as 
\begin{eqnarray}
&&\beta_{g_s}=-11+\frac{2}{3}(6+\delta)g_s^3, \quad  
\beta_g= -3g^3, 
\quad \beta_{g^\prime}=(7+4Y^2\delta)g^{\prime 3}, \nonumber \\ 
&&\beta_{h_t}=h_t\left(\frac{9}{2}h_t^2-8g_s^2-\frac{9}{4}g^2
-\frac{17}{12}g^{\prime 2}\right), \quad
\beta_{y_k}=y_k\left(y_k^2+3y_D^2+\sum_iy_i^2\right),\nonumber \\ 
&&\beta_{y_D}=y_D\left(-8g_3^3-6Y^2\delta g^{\prime 2}+4y_D^2+\sum_iy_i^2
\right), \nonumber \\
&&\beta_{h_{1,2}}=h_{1,2}\left[-\frac{9}{4}g^2-
\frac{3}{4}g^{\prime 2}+5(h_1^2+h_2^2)+3h_3^2+\frac{1}{2}\sum_iy_i^2\right],
\nonumber \\
&&\beta_{h_3}=h_3\left[-\frac{9}{4}g^2-
\frac{3}{4}g^{\prime 2}+2(h_1^2+h_2^2)+6h_3^2+\frac{1}{2}\sum_iy_i^2\right],
\end{eqnarray}
where $\delta$ stands for the number of extra color triplets $D_{L,R}$. 
Since $D_{L,R}$ is assumed to be light in this study, 
$\delta$ is treated as 1.
The Yukawa coupling $h_D$ with the ordinary quarks is assumed to be 
small enough and then its contribution is neglected in these equations.
  
\section*{Acknowledgements}
This work is partially supported by MEXT Grant-in-Aid 
for Scientific Research on Innovative Areas (Grant No. 26104009).

\newpage
\bibliographystyle{unsrt}

\end{document}